\documentclass[aps,prl,reprint,superscriptaddress]{revtex4-2}

\usepackage{graphicx}
\usepackage{dcolumn}
\usepackage{bm}
\usepackage{lineno}
\usepackage{ulem}
\usepackage{soul}
\usepackage{color}
\usepackage{array}
\usepackage{booktabs}
\usepackage{multirow, makecell}
\usepackage{physics}
\usepackage{lineno}
\usepackage[colorlinks=true, allcolors=blue]{hyperref}%

\newcommand{\UV}{DFT+$U$+$V$}

\usepackage{lipsum}
\makeatletter
\newcommand*{\balancecolsandclearpage}{%
  \close@column@grid
  \cleardoublepage
  \twocolumngrid
}

\begin{document}

\title{Intersite Coulomb Interactions in Charge Ordered Systems}

\author{Bo Gyu Jang}
\altaffiliation{Present Address: Theoretical Division, Los Alamos National Laboratory, Los Alamos, New Mexico 87545, USA}
\affiliation{Korea Institute for Advanced Study, Seoul 02455, Republic of Korea}

\author{Minjae Kim}
\email[Email: ]{garix.minjae.kim@gmail.com}
\affiliation{Korea Institute for Advanced Study, Seoul 02455, Republic of Korea}

\author{Sang-Hoon Lee}
\affiliation{Korea Institute for Advanced Study, Seoul 02455, Republic of Korea}

\author{Wooil Yang}
\affiliation{Department of Physics, Pohang University of Science and Technology, Pohang 37673, Republic of Korea}
\affiliation{Korea Institute for Advanced Study, Seoul 02455, Republic of Korea}

\author{Seung-Hoon Jhi}
\affiliation{Department of Physics, Pohang University of Science and Technology, Pohang 37673, Republic of Korea}

\author{Young-Woo Son}
\email[Email: ]{hand@kias.re.kr}
\affiliation{Korea Institute for Advanced Study, Seoul 02455, Republic of Korea}

\date{\today}

\begin{abstract}
Using {\it ab initio} approaches for extended Hubbard interactions coupled to phonons, we reveal that the intersite Coulomb interaction plays important roles in determining various distinctive phases of the paradigmatic charge ordered materials of Ba$_{1-x}$K$_x A$O$_3$ ($A=$ Bi and Sb). We demonstrated that all their salient doping dependent experiment features such as breathing instabilities, anomalous phonon dispersions, and transition between charge-density wave and superconducting states can be accounted very well if self-consistently obtained nearest neighbor Hubbard interaction are included, thus
establishing a minimal criterion for reliable descriptions of spontaneous charge orders in solids. 
\end{abstract}

\maketitle


Since Verwey found the metal-to-insulator transition (MIT) 
in magnetite (Fe$_{3}$O$_{4}$) owing to
long-range order of alternating Fe$^{2+}$ and Fe$^{3+}$ ions \cite{Verwey1939}, the charge ordered state has been one of the central issues in condensed matter physics. 
It often occurs near MITs, superconducting (SC), or charge density wave (CDW) states~\cite{Yamada1996, VanAken2003,Tranquada1995, Da2014,Da2015, Frano2020,Mattheiss1988, Cava1988}. 
These local charge ordering (CO) can lead to colossal magnetoresistance, 
ferroelectricity or multiferroicity~\cite{Ramirez1997,Van2008}. In addition, the CO in high-temperature cuprate superconductors has also been studied intensively to understand its roles as a leading competitor of SC state~\cite{Tranquada1995, Da2014, Da2015, Frano2020}.

Charge ordered materials host atoms with disparate charging states 
that are closely placed, 
invoking the strong Coulomb interactions and distorting lattices to relieve their energetic cost~\cite{Attfield2006}.
Hence, theories for the systems should treat the interaction as well as its coupling to lattices on an equal footing.
Typical approaches based on density functional theory with the local density approximation (DFT-LDA)~\cite{LDA} or generalized gradient approximation (GGA)~\cite{GGA} fail to describe their properties~\cite{Anisimov1996, Tsumuraya2020}. The addition of on-site Hubbard interaction ($U$) within DFT (DFT$+U$)~\cite{Hubbard1963, Schuler2013} captures a correct CO state when the intersite interaction is screened~\cite{Anisimov1996}.
Except few cases, however, it is insufficient in obtaining the ground states~\cite{Hitoshi2000, Yamamoto2007,Terletska2017}.
Therefore, advanced methods beyond the local corrections are needed to figure the role of the interactions in COs.

Perovskite BaBiO$_{3}$ (BBO) and BaSbO$_{3}$ (BSO)  
are prototypical CO materials that are not well understood by DFT-GGA and DFT+$U$~\cite{Cox1976, Cox1979, Kim2021,
Liechtenstein1991,Meregalli1998,Meregalli1999, Thonhauser2006, Korotin2012, Korotin2014, Franchini2010,Franchini2009, Yin2013, Li2019}. 
They show CDW states characterized by breathing distortion of oxygen octahedra with the charge disproportionations of Bi ions~\cite{Cox1976, Cox1979, Kim2021}. The CDW is suppressed by substituting Ba with K and SC phase occurs at a relatively higher transition temperature ($T_c$)~\cite{Mattheiss1988, Cava1988}. 
A few recent studies based on 
the Heyd-Scuseria-Ernzerhof (HSE) hybrid functional 
and $GW$ approximation (GWA)
can capture the insulating ground state of BBO and 
obtain enhanced electron-phonon ($e$-$ph$) interactions, providing a clue to understanding the observed $T_{c}$~\cite{Franchini2009, Franchini2010, Yin2013, Li2019, Zhihong2022}. However, realistic full phonon spectra to understand the transition between SC and CDW states
for experimentally 
accessible doping levels are hardly available due to demanding computational resources and computed frequency is usually overestimated for strongly coupled phonons~\cite{Meregalli1998, Yin2013, Li2019}.      

On the other hand, 
recent developments on self-consistent evaluations
of $U$~\cite{Agapito2015,Tancogne2018} and intersite Hubbard interactions ($V$)~\cite{Lee2020, Tancogne2020} within DFT (DFT+$U$+$V$)~\cite{Campo2010}  
successfully describe various properties of solids~\cite{Lee2020, Tancogne2020,Campo2010, Liu2020,Klik2011, Cococcioni2019, Ricca2020, Timrov2020PRB, Yang2021, Timrov2021PRB, Timrov2022arXiv,Yang2022JPCM}.
Owing to their low computational cost comparable to DFT-LDA and improved accuracy to GWA~\cite{Lee2020, Yang2021},
the new method provides an opportunity to study the correlated solids in large scale structures and full phase space of interests.
Motivated by these developments as well as the works on BBO and BSO, we have carried out {\it ab initio} study to explore the role of the interactions for interplay between their electronic and structural properties.

In this Letter, 
we theoretically demonstrate that the evolution of charge ordered states in bismuthates and antimonates with potassium doping is essentially controlled by the doping-dependent non-local Coulomb interaction. 
Our new parameter-free {\it ab initio} method for extended Hubbard interactions~\cite{Lee2020, Yang2021} can compute electronic energy bands as well as full phonon dispersions of Ba$_{1-x}$K$_x A$O$_3$ ($A=$ Bi and Sb) for the whole phase space with structural phase transitions, agreeing well with all the key measurements only when intersite Coulomb interactions are included. 
Our establishments
imply that the explicit treatment of the non-local interactions is critical for the description of the intertwined charge and lattice degrees of freedom in charge ordered materials.

Our DFT+$U$+$V$ method uses the total energy functional of $E_\text{tot}=E_\text{DFT}+E_\text{Hub}$ that
can be decomposed into (semi)local density functional of $E_\textrm{DFT}$ and
Hubbard functional with double counting corrections~\cite{Campo2010}, 
$
E_\textrm{Hub} 
=\frac{1}{2}\sum_I U_I \sum_{m,m',\sigma}  (\delta_{mm'}-n^{II\sigma}_{mm'})n^{II\sigma}_{m'm}
$
$
-\frac{1}{2}\sum_{\{I,J\}}V_{IJ}\sum_{m,m',\sigma}n^{IJ\sigma}_{mm'}n^{JI\sigma}_{m'm},
$
where the generalized occupation matrix is $n_{mm'}^{IJ \sigma} = \sum_{{\bf k}\nu}f_{{\bf k}\nu}^{\sigma}$ $\bra{\psi_{{\bf k}\nu}^{\sigma}}\ket{\phi_{m'}^{J}}$ $\bra{\phi_{m}^{I}}\ket{\psi_{{\bf k}\nu}^{\sigma}}$, $f_{{\bf k}\nu}^{\sigma}$ the Fermi-Dirac function of Kohn-Sham orbital of $\psi_{{\bf k}\nu}^{\sigma}$ of the $\nu$th band with spin $\sigma$ at momentum ${\bf k}$, and $\phi_{m}^{I}$ the localized orbitals with angular quantum number $m$. 
Here, 
$I$ and $J$ are abbreviated indexes for atomic positions and principal and azimuthal quantum numbers together and $\{I,J\}$ denotes a pair of atoms within the nearest neighboring distances. 
We obtain self-consistent $U_I$ and $V_{IJ}$ using new pseudohybrid functionals for Hubbard interactions~\cite{Agapito2015, Tancogne2018,Lee2020, Tancogne2020}. We also consider rotationally invariant interactions~\cite{Campo2010, Lee2020} so that $U$ and $V$ for valence $s$ and $p$ orbitals of Bi (Sb) and 2$p$ orbital of O are computed as shown in Table~\ref{tab:table1}. Detailed  parameters are in Supplementary Materials (SM)~\cite{suppl}.

\begin{table}[b]
	\caption{\label{tab:table1}Calculated $U$ and $V$ (in eV) for Ba$A$O$_3$ ($A =$ Bi and Sb). $U_{s(p)}^{A}$ ($U_p^O)$ is on-site Hubbard parameters of valence $s(p)$ orbital of $A(O)$ atom. $V_{sp(pp)}$ is the intersite Hubbard parameters between $s(p)$ orbital of $A$ and $p$ orbital of O. }
	\begin{ruledtabular}
		\begin{tabular}{cccccc}
			$A$&$U_{s}^A$&$U_{p}^A$&$U_p^O$&$V_{sp}$&$V_{pp}$\\ \hline
			Bi&1.03&0.11&8.18&1.78&1.59\\
			Sb&0.92&0.13&8.18&1.86&1.61\\
		\end{tabular}
	\end{ruledtabular}
\end{table}

\begin{figure}[t] 
	\includegraphics[width=1\linewidth]{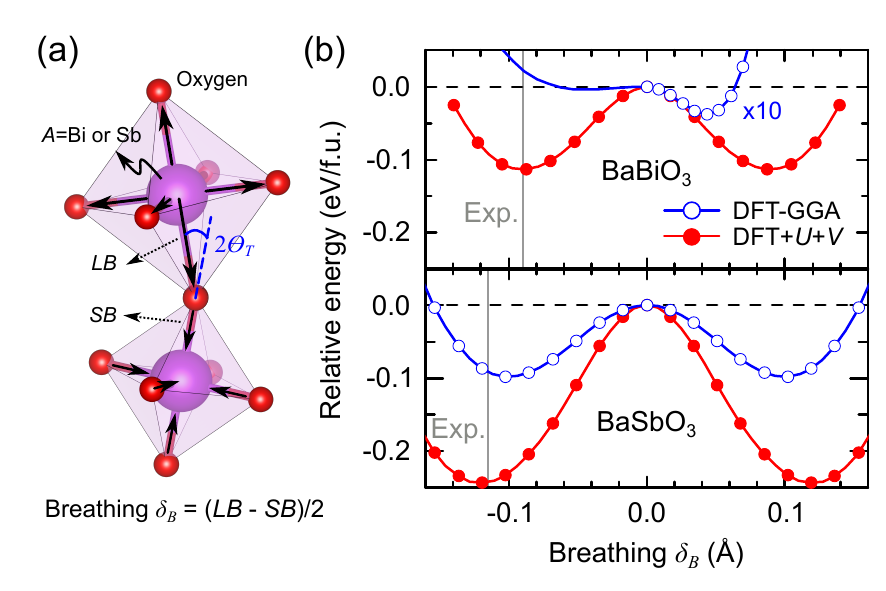}%
	\caption{(a) Atomic model for distorted octahedra in CDW state of BB(S)O. Breathing distortion of $\delta_B$ and tilting angle of $\theta_T$ are defined as an averaged length difference between Bi(Sb)-O bonds and as an angle between vertical axes belong to adjacent octahedra, respectively. (b) Double well potential as a function of $\delta_B$. Blue, red, and grey lines indicate DFT-GGA, \UV\, and experiments~\cite{Pei1990, Kim2021}, respectively. The DFT-GGA results for $\delta_B>0$ are enlarged by ten times to show the minimum for breathing distortion clearly. }
	\label{fig:Fig1}
\end{figure}

At low-temperature, BBO is the monoclinic structure having oxygen octahedra with breathing distortion of displacement by $\delta_B$ and a tilting angle of $\theta_T$ between them as shown in Fig. \ref{fig:Fig1}(a), while BSO  has $\delta_B$ only, resulting in fcc structure (Fm$\bar{3}$m)~\cite{Kim2021}. 
We first investigate the artificial double-well potential induced by $\delta_B$ without tilting [Fig.\ref{fig:Fig1}(b)]. 
For BBO, our DFT-GGA calculation underestimates $\delta_B$ with a very shallow potential well~\cite{Meregalli1998,Thonhauser2006}. The breathing distortion, however, is significantly enhanced in \UV calculation.
Unlike BBO, BSO has a relatively deep potential with DFT-GGA. Nonetheless, the energy gain from the breathing distortion in BSO also becomes larger 
with the extended Hubbard interactions.

\begin{table}[b]
	\caption{\label{tab:table2}Calculated structural and electronic properties of BBO along with computational and experimental data from previous studies. $v$: volume, $\beta$: monoclinic angle, $\delta_B$: breathing distortion, $\theta_T$: tilting distortion, and $E_{g}$: band gap  }
	\begin{ruledtabular}
		\begin{tabular}{ccccc}
			&Experiments
			\footnote{References \onlinecite{Cox1976,Pei1990,Kennedy2006} }
			&HSE\footnote{Reference \onlinecite{Franchini2010}}
			&DFT-GGA&\UV \\ 
			\hline
			$v$ (\AA$^{3}$)&81.80 $\sim$ 82.54
			&82.10&85.03&82.94\\
			$\beta$ (deg)&90.16 $\sim$ 90.27 &90.24&90.39&90.34\\
			$\delta_B$ (\AA)&0.08 $\sim$ 0.09 &0.09&0.08&0.10\\
			$\theta_T$ ($^\circ$)&10.12 $\sim$ 10.72 &11.9&11.75&10.46\\
			$E_{g}$ (eV)&0.8 $\sim$ 1.1&0.84&0.0&0.99
		\end{tabular}
	\end{ruledtabular}
\end{table}

As shown in the model potential wells in Fig.~\ref{fig:Fig1} (b), 
\UV\ well captures a part of CDW states of the undoped cases. Our results on a fully relaxed monoclinic BBO are summarized and compared with the previous studies (Table~\ref{tab:table2}). If tilting of $\theta_T$ is allowed, DFT-GGA accidentally reproduces $\delta_B$ owing to its overestimation of volume by $\sim$ 4\%\ and $\theta_T$ by $\sim$ 14\%  but still cannot describe the insulating gap of CDW state like a previous work~\cite{Franchini2010}. On the other hand, \UV\ calculation well describes all the critical experimental parameters.

\begin{figure}
	\includegraphics[width=1\linewidth]{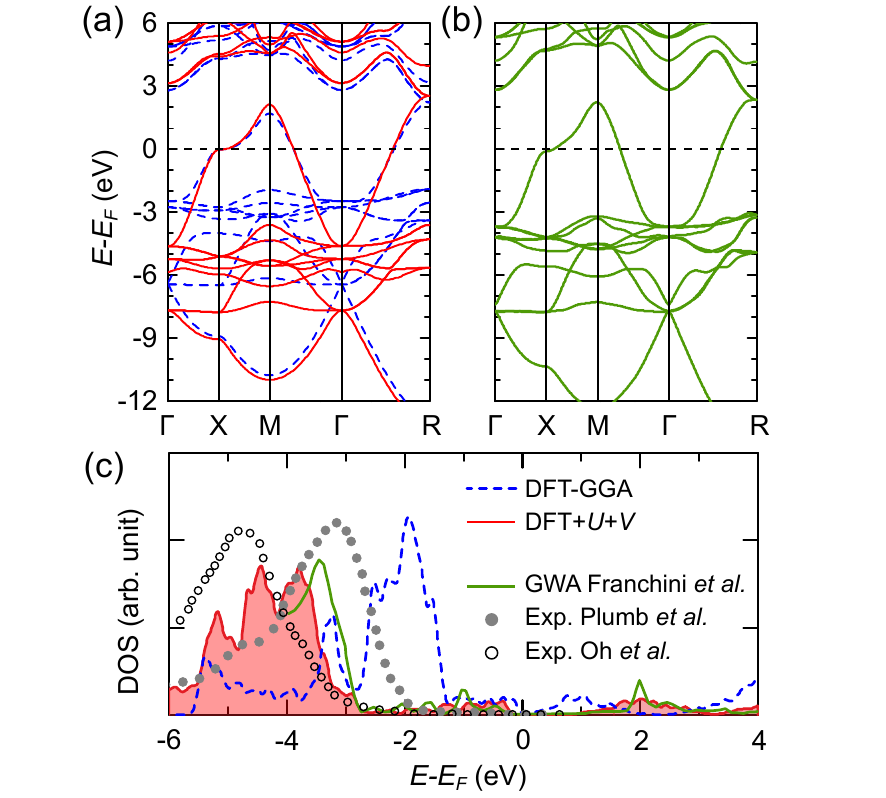}%
	\caption{Energy bands of BBO of cubic perovskite structure obtained from (a) DFT (blue) and \UV\ (red) and (b) GWA method. Here we use $GW_0$ approximation for the GWA. (c) Density of states for monoclinic CDW structure. Blue, red, and green line indicate DFT, \UV\, and previous GWA results~\cite{Franchini2010}, respectively. Empty circles are from photoemission spectroscopy measurements~\cite{Plumb2016, Oh2018}. } 
	\label{fig:Fig2}
\end{figure}

To investigate the effects of $U$ and $V$ in metallic BBO above the CDW transition temperature of 800 K,
we compute the energy bands of BBO in the perfect cubic perovskite phase with experimental volume, using DFT-GGA, \UV, and GWA as shown in Figs.~\ref{fig:Fig2} (a) and (b).  The width of the energy band crossing the Fermi level ($E_{F}$) is enhanced with both \UV and GWA compared with one with DFT-GGA. 
We also note that the DFT-GGA bands for fully occupied states associated with oxygen $p$ orbitals 
shifted down by including $U^O_p$ in the \UV~and by the self-energy corrections in the GWA method.

Effects of intersite interaction are also conspicuous for electronic structures of the CDW state. Figure~\ref{fig:Fig2} (c) shows the calculated density of states (DOS) of BBO with the fully relaxed monoclinic CDW structure obtained from each method (Table \ref{tab:table2}) together with previous GWA results and photoemission spectroscopy (PES) data~\cite{Plumb2016, Oh2018}. The two experiments show the quite different positions of the highest PES peak below $E_F$ ($-3$ eV and $-5$ eV, respectively), that may originate from the different substrate conditions~\cite{Oh2018}. 
As shown in Fig.~\ref{fig:Fig2}(c), the DOS peak position for oxygen $2p$ orbitals from DFT-GGA is significantly off the experimental one. Furthermore, the energy gap of CDW state is absent despite the correct $\delta_B$. Unlike DFT-GGA results, the DOS using \UV\ well agrees with the experiments and the previous GWA calculation \cite{Franchini2010}. The O 2$p$ peak position is located at around $-4$ eV, which is in between two experimental results. The band gap is about 1 eV, consistent with the experimental results. 
Although $U^O_p$ is critical for the 
down shifted bands for O 2$p$ orbitals, 
we note that the DFT+$U$ without $V$ 
still substantially underestimates the CDW band gap 
(Table S1 in the SM).

The results so far demonstrate that our new method describes correctly the electronic and structural properties of undoped bismuthate and antimonates without serious computational costs. Thus, it enables the study of the non-local interaction effects on phonons with various K doping levels thoroughly, which could not be done with HSE or GWA method easily. From now on, we present the comprehensive phonon dispersions using the frozen phonon techniques~\cite{phonopy} with extended Hubbard interactions to examine the structural instability of potassium doped Ba$_{1-x}$K$_x$BiO$_3$ and Ba$_{1-x}$K$_x$SbO$_3$ (BKBO and BKSO). Related computational details are in SM~\cite{suppl}.

\begin{figure}[t]
	\includegraphics[width=1\linewidth]{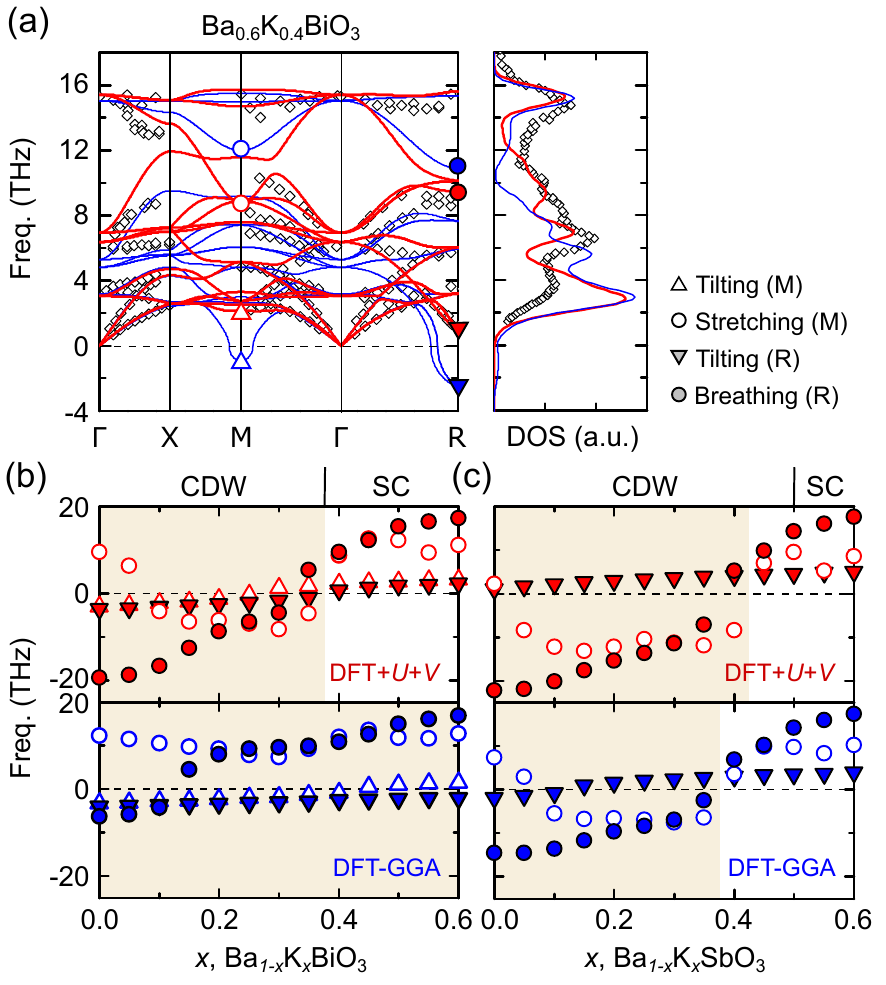}%
	\caption{(a) Phonon dispersion and DOS of Ba$_{0.6}$K$_{0.4}$BiO$_{3}$. Open diamonds and blue (red) lines indicate experiments \cite{Braden1995,Loong1992} and the DFT-GGA (\UV) results, respectively. Frequencies of the four selected modes are marked by open and filled triangles and circles at $M$ and $R$ points, respectively. Phonon frequency variations of the four modes as increasing K doping level in (b) BKBO and (c) BKSO. Top and bottom panels are for the modes obtained by \UV\ and DFT-GGA methods, respectively. Experimental~\cite{Pei1990, Kim2021} and our theoretical phase boundaries between CDW and SC states are shown on the upper abscissa and denoted by background color changes in (b) and (c), respectively.} 
	\label{fig:Fig3} 
\end{figure}

Figure~\ref{fig:Fig3} (a) shows the phonon dispersion and DOS of Ba$_{0.6}$K$_{0.4}$BiO$_{3}$ together with experimental data~\cite{Braden1995,Loong1992}. Here we focus four representative phonon modes related with CDW states in BKBO and BKSO systems; stretching and in-phase tilting modes at $M$ point and breathing and anti-phase titling modes at $R$ point. For stretching mode, Bi-O bond length changes along only two axes of oxgen octahedron while it changes along all three axes in the breathing mode. 
In Fig.~\ref{fig:Fig3}(a), it is immediately noticeable that the unstable in(anti)-phase tilting mode at M(R) point obtained by DFT-GGA hardens enough to be stable when $V$ is included, agreeing with experiments. Not only low frequency modes, but the high frequency optical branches obtained with $V$ also agree with experiments. Specifically, our result well matches the anomalous dispersion of LO mode along $\Gamma$X~\cite{Braden1995, Braden1996, Braden2002} related with the instability toward the charge ordering~\cite{Braden1996, Braden2002}.

We present doping-dependent evolution of frequencies for the selected phonon modes of BKBO and BSBO in Figs.~\ref{fig:Fig3}(b) and (c). 
Our calculations with the extended Hubbard interactions fruitfully reflect the measured trends of structural distortions for the both systems. 
In the case of BKBO shown in Fig.~\ref{fig:Fig3}(b),
the breathing mode computed using DFT-GGA becomes to be stable when $x\ge 0.15$ while it does only when $x\ge 0.4$ in the experiments \cite{Tajima1992, Braden2000, Braden2002} and our results with $V$. 
BKBO with $x\ge 0.4$ show SC states without structural distortions \cite{Cava1988, Fleming1988, Pei1990, Braden1995} while the tilting instabilities still remain in DFT-GGA as shown in Figs.~\ref{fig:Fig3} (a) and (b)~\cite{Li2019}.
This is in sharp contrast to complete absence of  instability in \UV results when $x\ge 0.4$.
Thus, the non-local Coulomb interaction is decisive in capturing the transition between CDW and SC states of BKBO. 

As already shown in Fig.~\ref{fig:Fig1} (b), the effect of $V$ is not as crucial in BKSO as it is in BKBO. Nevertheless, our method makes improvements in describing experimental phase diagram as shown in Fig.~\ref{fig:Fig3} (c). For undoped case, DFT-GGA shows the tilting instability in addition to the breathing instability while only the latter is observed in the experiment. Our calculation with $V$ results in a perfect fcc structure~\cite{Kim2021}. In addition, CDW phase survives up to higher doping of $x\simeq 0.4$ in \UV\ calculation, which is closer to the experimental phase boundary of $x\simeq 0.5$, due to the enhanced stretching instability.
Comprehensive comparisons between results from DFT+$U$ and DFT+$U$+$V$ 
and detailed discussions are in Sec. D, Figs. S2, S5 and S7 of the SM~\cite{suppl}.

\begin{figure} [t]
	\includegraphics[width=1\linewidth]{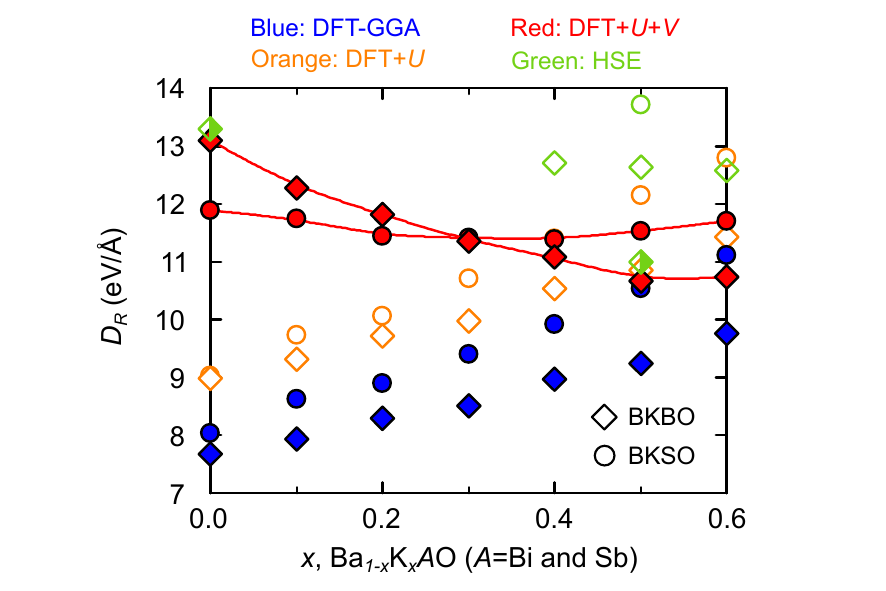}%
	\caption{
	Reduced $e$-$ph$ matrix element ($D^L_R$) as a function of doping $x$. Diamonds (circles) denote BKB(S)O and blue, orange, and red color indicate DFT-GGA, DFT+$U$, and \UV results, respectively. Open and right-half diamonds are HSE results for the element from Ref.~\onlinecite{Zhihong2022} and Ref.~\onlinecite{Yin2013}, respectively.
	} 
	\label{fig:Fig4}
\end{figure}

Finally, we estimate the effects of non-local interactions on $e$-$ph$ coupling constant ($\lambda$) for the breathing mode that is an important factor for the SC state.
Instead of calculating $\lambda$ explicitly, we compute `reduced $e$-$ph$ matrix element'~\cite{Yin2013,Korotin2014, Kang2019},
$D^{\mu{\bf k}}_{\alpha{\bf q}}=\partial \epsilon_{\mu{\bf k}}/{\partial {\bf u}^\alpha_{\bf q}}$
to compare with previous studies based on HSE directly~\cite{Yin2013, Zhihong2022}. 
Here, ${\bf u}^\alpha_{\bf q}$ is a displacement vector of phonon mode $\alpha$ with the wave-vector \textbf{q} and 
$\epsilon_{\mu \bf k}$
energy of $\mu$th band at ${\bf k}$.
Specifically, following previous works~\cite{Yin2013, Zhihong2022}, we obtain the reduced element of $D^L_{R}$ for the energy band crossing the $E_F$ at ${\bf k}$ of L point by
the breathing mode at ${\bf q}$ of R point (see Fig. S3 in SM~\cite{suppl}). 

For both systems, $D^L_R$ increases with doping within DFT-GGA and DFT+$U$ as shown in Fig.~\ref{fig:Fig4}. In sharp contrast to this, the matrix element with $V$ decreases with increasing $x$ for BKBO while it remains more or less the same for BKSO. 
So, the matrix element of the latter becomes larger than that of the former for $x>0.3$. 
Considering that the coupling can
be roughly estimated as $\lambda\sim(D^{\mu{\bf k}}_{\alpha{\bf q}}/\omega^\alpha_{\bf q})^2$
for phonon frequency $\omega_{\bf q}^\alpha$ 
~\cite{Yin2013,Korotin2014, Kang2019}, 
Fig.~\ref{fig:Fig4} implies a higher $T_c$ of BKSO than BKBO at larger doping levels with $V$, agreeing with a recent experiment~\cite{Kim2021}.
We also note that our estimations are consistent with two previous HSE results for a few selected dopings~\cite{Zhihong2022, Yin2013}.

The opposite trends of $D^L_R$'s for different computational methods can be understood by considering intersite interactions under lattice distortions. 
The paired octahedra with the breathing modes in Fig.~\ref{fig:Fig1} have an elongated long bond (LB) and a shrinking short bond (SB) between Bi(Sb) and oxygens~\cite{Kim2021}.
Based on a perturbation theory, $D^L_R$ can be written as 
$
D^L_R\delta u_B\simeq \varepsilon_\textrm{SB}-\varepsilon_\textrm{LB}-V(n_\textrm{SB}-n_\textrm{LB})
$
where $\varepsilon_\textrm{SB(LB)}$ and $n_\textrm{SB(LB)}$ are energy level and density matrix for SB(LB), respectively,
and $\delta u_B$ is the perturbation amplitude of the breathing mode
(See SM~\cite{suppl} for detailed derivations).
Since $n_\textrm{LB}>n_\textrm{SB}$ for lower dopings, the intersite interaction should enhance the matrix elements over those from DFT-GGA and DFT+$U$. As doping level is increased, the difference 
between $n_\textrm{LB}$ and $n_\textrm{SB}$ diminishes so that the effect of nonlocal interactions vanishes as shown in Fig.~\ref{fig:Fig4},
thus highlighting again a critical role of $V$ for doping dependent electron-phonon interactions. 

In summary, we present a comprehensive study on the doping dependent electronic and structural properties of prototypical charge ordered materials, BKBO and BKSO using a newly developed {\it ab initio} computational method. We demonstrated that the non-local Coulomb interactions between the nearest neighbours are essential physical parameters in determining the doping dependent evolution of CDW and SC states, highlighting nontrivial relationships between the nonlocal interaction, charge order and lattice distortion in correlated materials.   

\begin{acknowledgements}
B.G.J. M. K. and W. I. were supported by KIAS individual Grants (No. QP081301, CG083501, QP090101).
S.-H.J. was supported by NRF of Korea (Grant No. 2022R1A2C1006530) funded by the Korea government (MSIT).
Y.-W.S. was supported by NRF of Korea (Grant No. 2017R1A5A1014862, SRC program: vdWMRC center) and KIAS individual Grant (No. CG031509).
Computations were supported by the CAC of KIAS.
\end{acknowledgements}


%

\balancecolsandclearpage

\setcounter{table}{0}
\renewcommand{\thetable}{S\arabic{table}}%
\setcounter{figure}{0}
\renewcommand{\thefigure}{S\arabic{figure}}%
\setcounter{equation}{0}
\renewcommand{\theequation}{S\arabic{equation}}%

\section*{Supplementary Information}
\subsection{A. Pseudohybrid density functional for extended Hubbard interactions}

\begin{figure}[b]
	\includegraphics[width=1.0\linewidth]{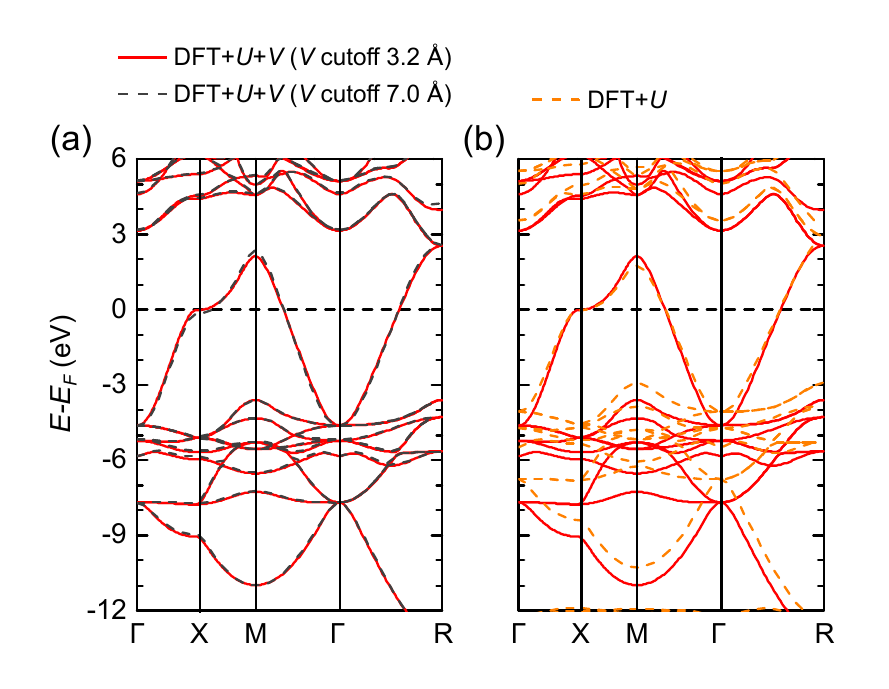}%
	\caption{(a) Energy bands of BBO of cubic perovskite structure obtained from \UV\ with different $V$ cutoff length. For 3.2 \AA cutoff case, only the nearest Bi-O pairs are considered while up to 3rd nearest Bi-O pairs are considered for 7.0 \AA cutoff case. (b) Comparison between energy band obtained from DFT+$U$ ($U^{Bi}_s, U^{Bi}_p$, and $U^{O}_{p}$) and \UV. Here, $U^{A}_{o}$ for $o$-orbital of atom $A$ shown in Table~\ref{T2} } 
	\label{fig:FigS2}
\end{figure}

The total energy formula with the extended Hubbard interactions including both on-site and intersite Coulomb interaction ($U$ and $V$) is given by \cite{Campo2010}
\begin{eqnarray}\label{e_dft}
	E_\textrm{DFT+U+V}=E_\textrm{DFT}+E_\textrm{UV},
\end{eqnarray}
where $E_\textrm{DFT}$ is the (semi)local density functional and $E_\textrm{UV}$ the extended Hubbard funcitonal. Within a rotationally invariant form with the fully localized limit (FLL) double counting correction, $E_\textrm{UV}$ is written as \cite{Campo2010}
\begin{eqnarray}\label{e_uv}
	E_\textrm{UV}&=&\sum_{I}\sum_{mm'\sigma}\frac{U^I}{2}(\delta_{mm'}-n^{II\sigma}_{mm'})n^{II\sigma}_{m'm} \nonumber\\
	& &-\sum_{\{I,J\}}\sum_{mm'\sigma}\frac{V^{IJ}}{2}n^{IJ\sigma}_{mm'}n^{JI\sigma}_{m'm},
\end{eqnarray}
where the generalized occupation matrix is $n_{mm'}^{IJ \sigma} = \sum_{{\bf k}\nu}f_{{\bf k}\nu}^{\sigma}\bra{\psi_{{\bf k}\nu}^{\sigma}}\ket{\phi_{m'}^{J}}\bra{\phi_{m}^{I}}\ket{\psi_{{\bf k}\nu}^{\sigma}}$, $f_{{\bf k}\nu}^{\sigma}$ the Fermi-Dirac function of Kohn-Sham orbital of $\psi_{{\bf k}\nu}^{\sigma}$ of the $\nu$th band with spin $\sigma$ at momentum ${\bf k}$. 
{The L\"{o}wdin orthonormalized atomic wave function, $\phi_{m}^{I}$ is used as a projector for the localized atomic orbital where $m$ is angular quatum number.}
Here, 
$I$ and $J$ are abbreviated indexes for atomic positions and principal and azimuthal quantum numbers together. 
$\{I,J\}$ in Eq.~\ref{e_uv} denotes a pair of atoms within the nearest neighboring distances.

\squeezetable
\begin{table}[t]
	\caption{Calculated structural and electronic properties of BBO along with computational and experimental data from previous studies. $v$: volume, $\beta$: monoclinic angle, $\delta_B$: breathing distortion, $\theta_T$: tilting distortion, and $E_{g}$: band gap  }{\label{T1}}
	\begin{ruledtabular}
		\begin{tabular}{ccccc}
			&DFT-GGA & DFT+$U$ (w/o O $2p$) & DFT+$U$ &\UV \\ 
			\hline
			$v$ (\AA$^{3}$)       & 85.03 & 85.03 & 82.19 & 82.94\\
			$\beta$ (deg)         & 90.39 & 90.40 & 90.27 & 90.34\\
			$\delta_B$ (\AA)      & 0.08  & 0.08  & 0.10 & 0.10\\
			$\theta_T$ ($^\circ$) & 11.75 & 11.76 & 9.71 & 10.46\\
			$E_{g}$ (eV)          & 0.00   & 0.00  & 0.71 & 0.99\\
			$U^{Bi}_{p}$          & -     & 0.14  & 0.09 & 0.11 \\
			$U^{O}_{p}$           & -     &  -    & 8.38 & 8.23 \\
		\end{tabular}
	\end{ruledtabular}
\end{table}

\subsection{B. Calculation details}

Following the ansatz by Mosey $et$ $al$.~\cite{Mosey2007PRB, Mosey2008JCP}, we extend the ACBN0 pseudohybrid functional for $U$~\cite{Agapito2015} into $V$~\cite{Lee2020} 
by considering the renormalized occupation number $N^{IJ\sigma}_{\psi_{\boldsymbol{k}v}}$ and density matrix $P^{IJ\sigma}_{mm'}$ for a pair of different atoms $I$ and $J$:
\begin{eqnarray}\label{eq:n}
	N^{IJ\sigma}_{\psi_{{\bf k}v}}&=&
	\sum_{{I}}\sum_{m}\bra{\psi_{{\bf k}v}^{\sigma}}\ket{\phi_{m}^{I}}\bra{\phi_{m}^{I}}\ket{\psi_{{\bf k}v}^{\sigma}} \nonumber \\
	& &+\sum_{{J}}\sum_{m'}\bra{\psi_{{\bf k}v}^{\sigma}}\ket{\phi_{m'}^{J}}\bra{\phi_{m'}^{J}}\ket{\psi_{{\bf k}v}^{\sigma}},\\
\label{eq:p}
	P^{IJ\sigma}_{mm'}&=&
	\sum_{{\bf k}v}
	w_{{\bf k}}
	f_{{\bf k}v}
	N^{IJ\sigma}_{\psi_{{\bf k}v}}
	\bra{\psi_{{\bf k}v}^{\sigma}}\ket{\phi_{m}^{I}}\bra{\phi_{m'}^{J}}\ket{\psi_{{\bf k}v}^{\sigma}},
\end{eqnarray}
where $w_{\boldsymbol{k}}$ is $\bf{k}$-point weight and $f_{{\bf k}v}$ is the Fermi-Dirac function of the Bloch state $\ket{\psi_{{\bf k}v}^{\sigma}}$.
By using Eq.~\ref{eq:n} and~\ref{eq:p}, $U^{I}$, $J^{I}$, and $V^{IJ}$ can be defined as~\cite{Agapito2015, Lee2020, Tancogne2020}:
\begin{equation}
\label{eq:u}\scriptstyle
	U^{I}=\frac{\sum_{m,\cdots,m'''} \sum_{\sigma\sigma'} P^{II\sigma}_{mm'}P^{II\sigma'}_{m''m'''} (mm'|m''m''')}
	{\sum_{m\neq m'} \sum_{\sigma} n^{II\sigma}_{mm} n^{II\sigma}_{m'm'} + \sum_{\{m\}} \sum_{\sigma} n^{II\sigma}_{mm} n^{II-\sigma}_{m'm'}},
\end{equation}
\begin{equation}
\label{eq:j}\scriptstyle
	J^{I}=\frac{\sum_{m,\cdots,m'''} \sum_{\sigma} P^{II\sigma}_{mm'}P^{II\sigma}_{m''m'''} (mm'|m''m''')}
	{\sum_{m\neq m'} \sum_{\sigma} n^{II\sigma}_{mm} n^{II\sigma}_{m'm'}},
\end{equation}
\begin{equation}
\label{eq:v}\scriptstyle
	V^{IJ}=
	\frac{1}{2}
	\frac{\sum_{m,\cdots,m'''}\sum_{\sigma\sigma'} [P^{II\sigma}_{mm}P^{JJ\sigma'}_{m'm'} - \delta_{\sigma\sigma'} P^{IJ\sigma}_{mm'}P^{JI\sigma'}_{m'm}](mm'|m''m''')}
	{\sum_{\{m\}} \sum_{\sigma\sigma'} [n^{II\sigma}_{mm}n^{JJ\sigma'}_{m'm'} - \delta_{\sigma\sigma'} n^{IJ\sigma}_{mm'}n^{JI\sigma'}_{m'm}]},
\end{equation}
where $(mm'|m''m''') \equiv \int d\textbf{r}_{1}d\textbf{r}_{2} 
\phi_{m}^{I*}(\textbf{r}_{1})\phi_{m'}^{I}(\textbf{r}_{1})
|\textbf{r}_{1}-\textbf{r}_{2}|^{-1}
\phi_{m''}^{J*}(\textbf{r}_{2})\phi_{m'''}^{J}(\textbf{r}_{2})
$.

\begin{figure}[t] 
	\includegraphics[width=0.8\linewidth]{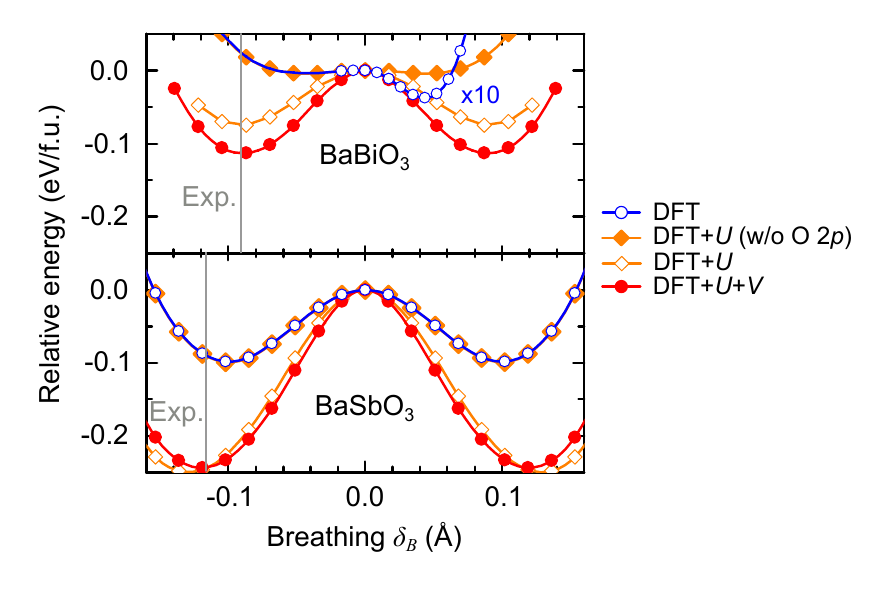}%
	\caption{Double well potential as a function of $\delta_B$. Blue, orange closed diamond, orange open diamond, and red indicate DFT-GGA, DFT+$U$ ($U^{Bi}_s, U^{Bi}_p$), DFT+$U$ ($U^{Bi}_s, U^{Bi}_p$, and $U^{O}_{p}$), and \UV\, respectively.}
	\label{fig:FigS1}
\end{figure}

We implemented the pseudohybrid density functionals for extended Hubbard interactions shown in Eqs.~\ref{eq:u},~\ref{eq:j}, and~\ref{eq:v} in  \texttt{QUANTUM ESPRESSO} package \cite{QE, Lee2020, Yang2021}. We used the norm-conserving pseudopotentials provided by the PseudoDojo project \cite{dojo}. The energy cutoff for charge density was set to 480 Ry. A 15$\times$15$\times$15 $k$-point mesh was used for self-consistent calculation of perfect perovskite structure (5 atoms in the unitcell). The $k$-mesh density was kept the same for all other calculations. The Hubbard $U$ and $V$ parameters are obtained during the self-consistent calculation from Eqs.~\ref{eq:u},~\ref{eq:j} and ~\ref{eq:v} as shown in Table \ref{T2}. 

\begin{figure}[b]
\includegraphics[width=1.0\linewidth]{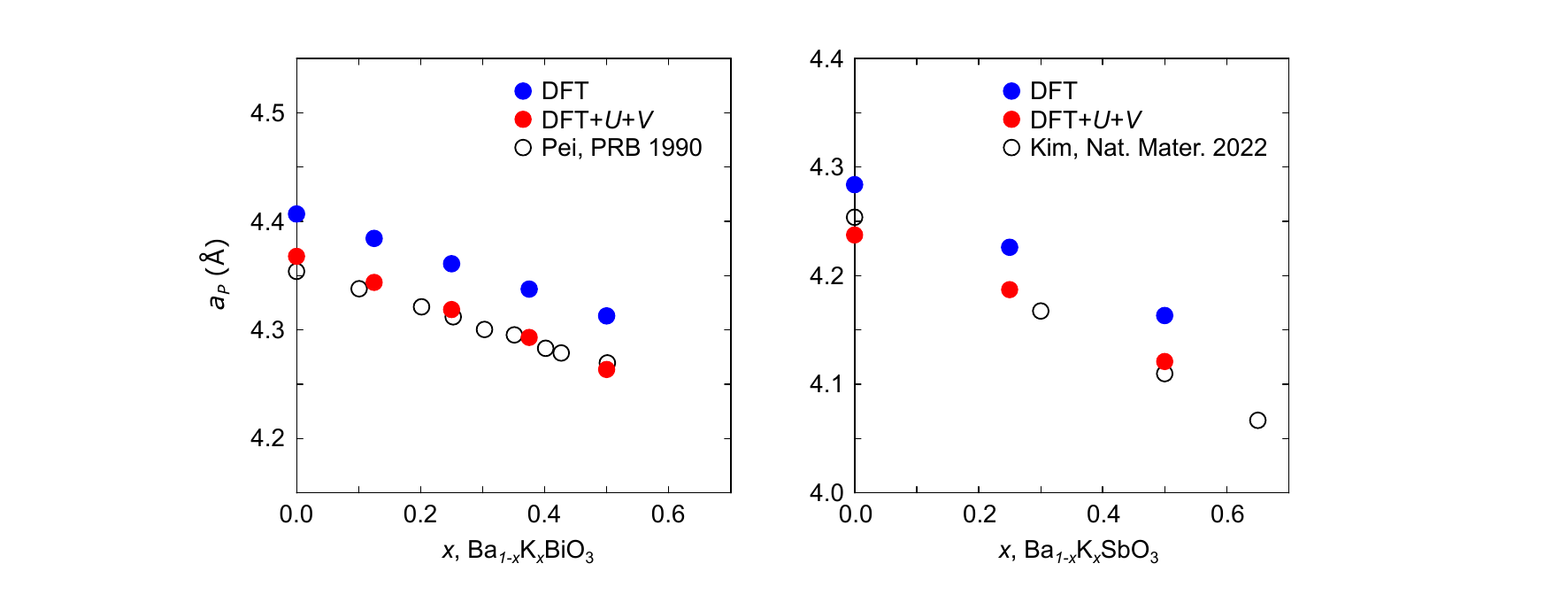}%
\caption{Experimental pseudocubic lattice parameter $a_{P}$ as a function of K doping of $x$ for BKBO and BKSO~\cite{Pei1990, Kim2021}. Cubic supercell structures up to 40 atoms are used to model realistic K substitution structures. DFT-GGA calculations overestimate the lattice constant while DFT+$U$+$V$ results well agree with experimental observations. \label{Fig.S.lattice}}
\end{figure}

\begin{figure}[t]
\includegraphics[width=1.0\linewidth]{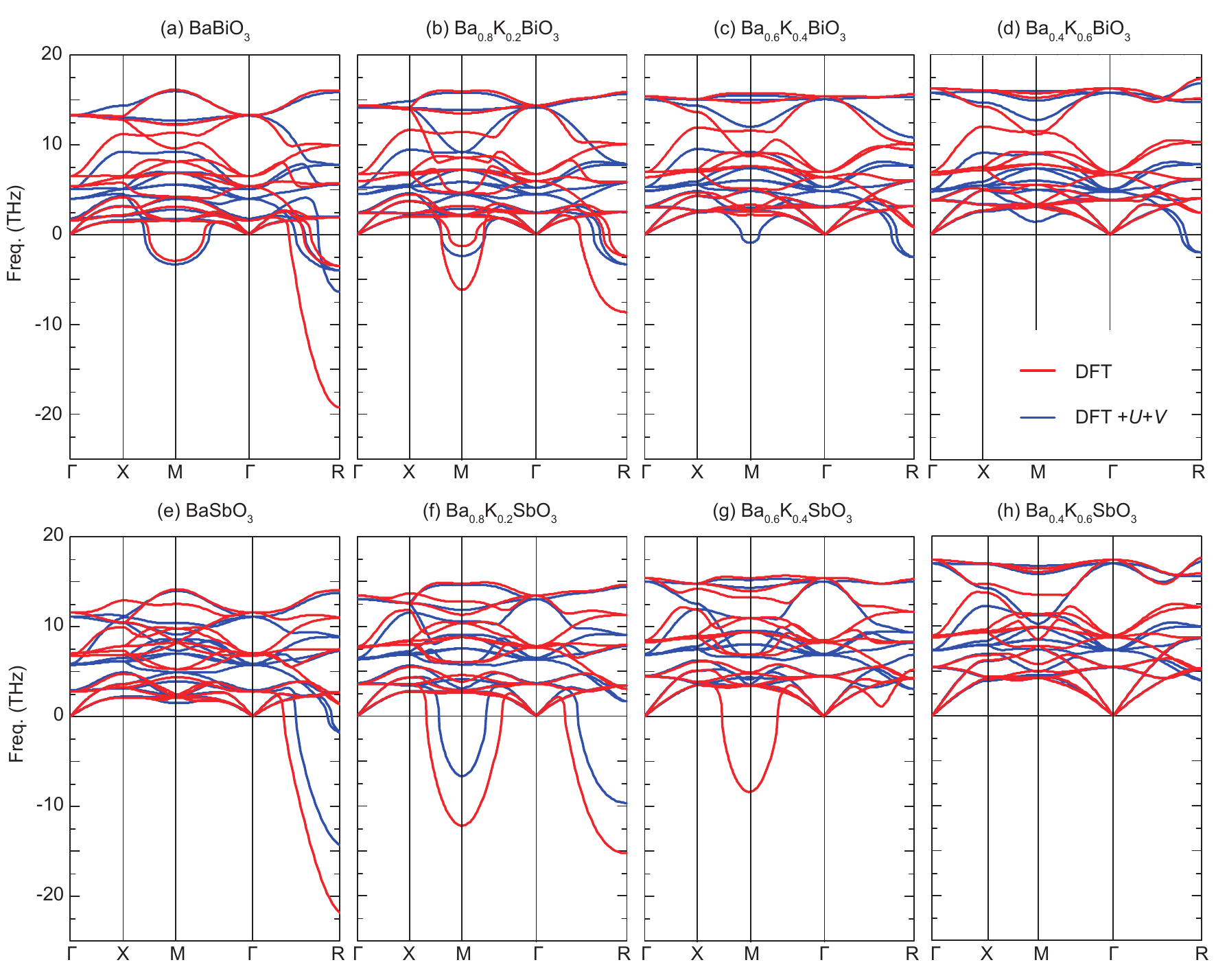}%
\caption{Phonon band structures of BKBO and BKSO. Blue and red lines indicate DFT and DFT+$U$+$V$ results, respectively. \label{Fig.S.phonon}}
\end{figure}

The potassium (K) substitution effects are simulated by reducing the number of electrons in the unitcell and inserting a compensating background charge. 
For lattice optimization with doping, we used large cubic supercell structures containing 40 atoms to model realistic doping. As shown in Fig.~\ref{Fig.S.lattice}, our optimized pseudocubic lattice parameter of $a_P$ using DFT+$U$+$V$ formalism
agrees with experimental data~\cite{Pei1990, Kim2021} very well while DFT-GGA fails.
With these optimized structures, 
the phonon dispersions were calculated 
using the finite displacement method by \texttt{Phonopy} \cite{phonopy}. 
{We used fixed $U$ and $V$ values, which are obtained from the equilibrium structure, to calculate the supercell structures with atomic displacement. }
Full phonon band structures of BKBO and BKSO are shown in Fig. \ref{Fig.S.phonon}.  

For the direct comparison with the previous HSE result (Fig.4 of main text), the reduced matrix element for estimating $e$-$ph$ interaction is obtained by using the same scheme used in Ref.~\onlinecite{Yin2013}. For example, the breathing distortion induces band splitting as shown Fig. \ref{Fig.S.split}. The energy splitting at L point ($\Delta E$) is used to determine the matrix element with the given oxygen displacement.

\begin{figure}[b]
	\includegraphics[width=1.0\linewidth]{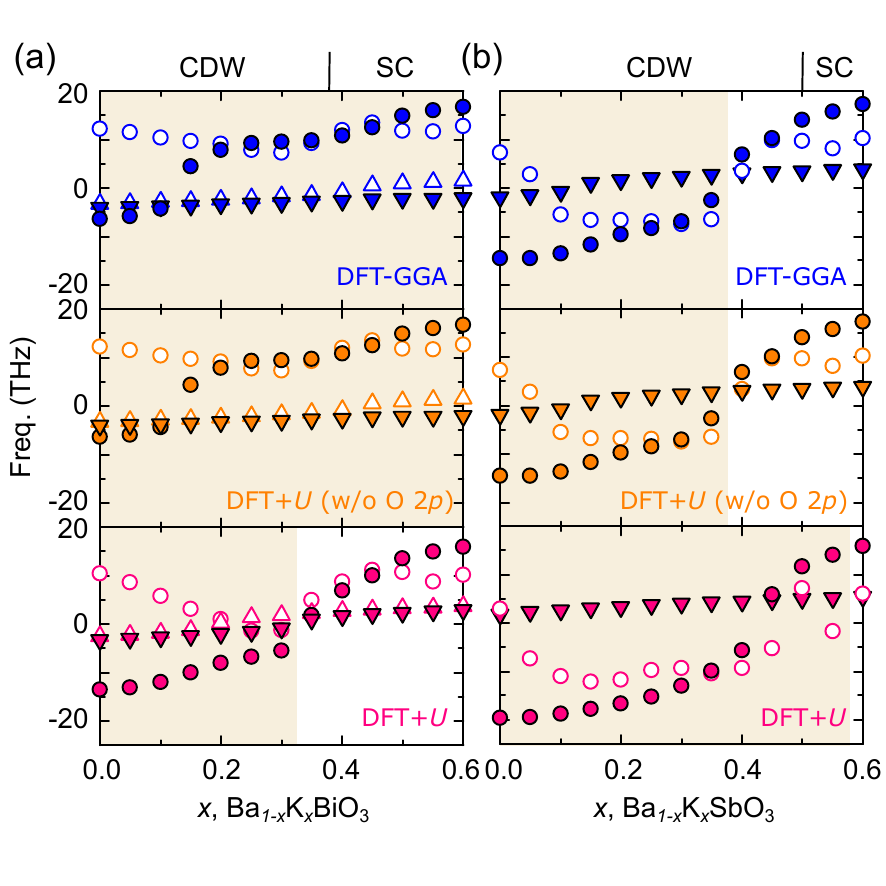}%
	\caption{Phonon frequency variations of the four modes as increasing K doping level in (a) BKBO and (b) BKSO. Top, middle, and bottom panels are for the modes obtained by DFT-GGA, DFT+$U$ ($U^{Bi}_s, U^{Bi}_p$), and DFT+$U$ ($U^{Bi}_s, U^{Bi}_p$, and $U^{O}_{p}$) methods, respectively. Experimental~\cite{Pei1990, Kim2021} and our theoretical phase boundaries between CDW and SC states are shown on the upper abscissa and denoted by background color changes in (a) and (b), respectively.} 
	\label{fig:FigS3} 
\end{figure}

\subsection{C. Doping dependence of the reduced {\textit{el-ph}} matrix elements}

\squeezetable
\begin{table}[b]
	\caption{Calculated $U$ and $V$ (in eV) for Ba$A_{1-x}$K$_{x}$O$_3$ ($A =$ Bi and Sb). $U_{s(p)}^{A}$ ($U_p^O)$ is on-site Hubbard parameters of valence $s(p)$ orbital of $A(O)$ atom. $V_{sp(pp)}$ is the inter-site Hubbard parameters between $s(p)$ orbital of $A$ and $p$ orbital of O. $V_{sp}^{Ba}$ is the inter-site Hubbard parameters of between $s$ orbital of $Ba$ and $p$ orbital of O. }{\label{T2}}
	\begin{ruledtabular}
		\begin{center}
			\begin{tabular}{c|ccccccc|cccccc}
			$x$&$U_{s}^{Bi}$&$U_{p}^{Bi}$&$U_p^O$&$V_{sp}$&$V_{pp}$&$V_{sp}^{Ba}$& &$U_{s}^{Sb}$&$U_{p}^{Sb}$&$U_p^O$&$V_{sp}$&$V_{pp}$&$V_{sp}^{Ba}$\\ \hline
			$0.00$&1.03&0.11&8.18&1.78&1.59&0.77&&0.92&0.13&8.18&1.86&1.61&0.75\\
			$0.05$&1.00&0.11&8.17&1.76&1.58&0.76&&0.95&0.13&8.15&1.82&1.59&0.74\\
			$0.10$&0.98&0.11&8.15&1.74&1.58&0.75&&0.91&0.13&8.12&1.78&1.58&0.74\\
			$0.15$&0.96&0.11&8.14&1.72&1.57&0.75&&0.88&0.13&8.08&1.74&1.56&0.73\\
			$0.20$&0.94&0.11&8.13&1.70&1.56&0.74&&0.85&0.13&8.05&1.71&1.55&0.72\\
			$0.25$&0.92&0.11&8.11&1.68&1.56&0.74&&0.83&0.12&8.01&1.68&1.54&0.71\\
			$0.30$&0.90&0.11&8.09&1.66&1.55&0.73&&0.80&0.12&7.97&1.65&1.52&0.71\\
			$0.35$&0.89&0.11&8.08&1.64&1.54&0.72&&0.78&0.12&7.93&1.63&1.51&0.69\\
			$0.40$&0.87&0.11&8.06&1.63&1.54&0.72&&0.76&0.12&7.89&1.60&1.49&0.68\\
			$0.45$&0.85&0.10&8.04&1.61&1.52&0.72&&0.74&0.12&7.84&1.58&1.48&0.68\\
			$0.50$&0.83&0.10&8.02&1.59&1.52&0.70&&0.72&0.12&7.79&1.55&1.46&0.67\\
			$0.55$&0.81&0.10&8.00&1.57&1.51&0.70&&0.70&0.12&7.74&1.53&1.44&0.66\\
			$0.60$&0.80&0.10&7.98&1.56&1.50&0.69&&0.68&0.12&7.69&1.51&1.43&0.65\\
			\end{tabular}
		\end{center}
	\end{ruledtabular}
\end{table}

\begin{figure}[t]
	\includegraphics[width=0.6\linewidth]{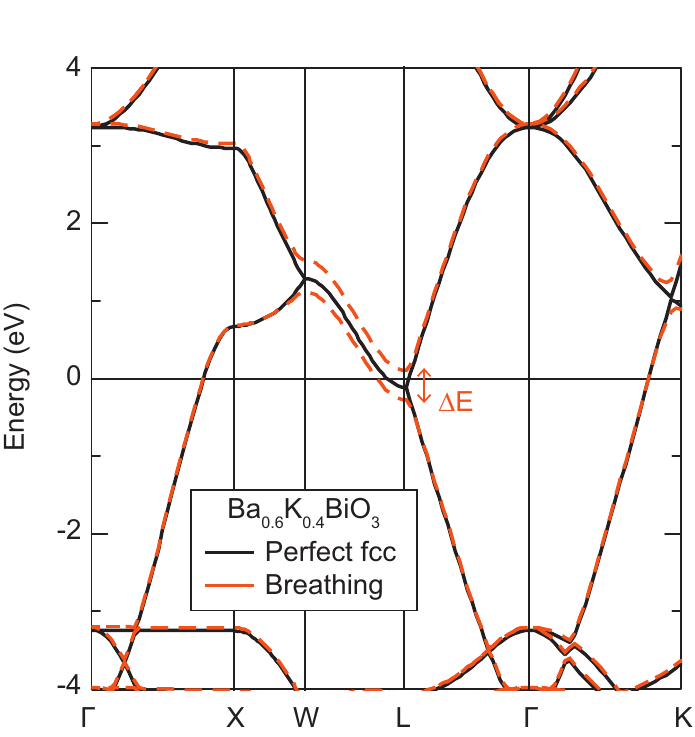}%
	\caption{Electronic band structure of Ba$_{0.6}$K$_{0.4}$BiO$_{3}$ with and with out the oxygen-breathing displacement obtained from DFT+$U$+$V$. The breathing distortion induces band splitting dipicted by orange dotted lines. The band splitting of $\Delta E$ indicated by the arrow are used to calculate the reduced $e$-$ph$ matrix element of $D^L_R$.} \label{Fig.S.split}
\end{figure}

The essential trend of the K doping ($x$) 
dependent variation of the reduced \textit{el-ph} matrix ($D^L_R$) of Ba$_{1-x}$K$_{x}$$A$O$_{3}$ 
($A$=Bi, Sb) shown in Fig. 4 of the main text can be understood by considering simplified perturbative expansions of interactions within DFT+$U$+$V$ scheme. The extended Hubbard interactions are treated as perturbations to the Kohn-Sham (KS) equation of (semi)local functional such that 
\begin{equation}\label{KS_with_UV}
{\mathcal H}_\textrm{total}={\mathcal H}_\textrm{DFT}+{\mathcal H}_\textrm{UV}
\end{equation}
where ${\mathcal H}_\textrm{total}\psi^\sigma_{{\bf k}\nu}= \epsilon^\sigma_{{\bf k}\nu} \psi^\sigma_{{\bf k}\nu}$ and 
${\mathcal H}_\textrm{DFT}\varphi^\sigma_{{\bf k}\nu}\equiv
	\left[-\nabla^{2}+V_\textrm{DFT}\right]\varphi^\sigma_{{\bf k}\nu}  = \varepsilon^\sigma_{{\bf k}\nu} \varphi^\sigma_{{\bf k}\nu}$.
Here, $V_\textrm{DFT}$ is the KS potential corresponding to (semi)local energy functional ($E_\textrm{DFT}$) and $\varphi^\sigma_{{\bf k}\nu}$ the KS wavefunction with eigenvalue of  $\varepsilon^\sigma_{{\bf k}\nu}$.
The perturbation can be written as 
${\mathcal H}_\textrm{UV} \psi^\sigma_{{\bf k}\nu}= {\delta E_\textrm{UV}}/
\delta
(\psi^{\sigma}_{{\bf k}\nu})^*$.
As discussed above, we only consider the alternation of energy bands crossing the Fermi level at the $L$ point in the presence of breathing distortion at the $R$ point as shown in Fig~\ref{Fig.S.split} so that the momentum, spin and band indexes will be dropped hereafter.

The energy band considered here mainly originate from the anti-bonding state between $s$ orbtial of Bi (Sb)
at the center of octahedra and $p$ orbitals of oxygen at their vertices~\cite{Kim2015}. Thus, the pair of distorted perovskites (see Fig. 1a of the main text) has elongated and shortened bonds for the anti-bonding states. We will call them the long and short bonding (LB and SB), respectively. From these considerations, the split energies of the band by the breathing mode in Fig.~\ref{Fig.S.split} can be approximated as energy eigenvalues of $\epsilon_\textrm{LB}$ and $\epsilon_\textrm{SB}$ (or $\varepsilon_\textrm{LB}$ and $\varepsilon_\textrm{SB}$ without ${\mathcal H}_\textrm{UV}$) for the static atomic configurations associated with LB and SB, respectively. The resulting difference of $\epsilon_\textrm{SB} - \epsilon_\textrm{LB}$ (or $\varepsilon_\textrm{SB}-\varepsilon_\textrm{LB}$ without ${\mathcal H}_\textrm{UV}$) can be assigned as $D^L_R$ because the energy level of the anti-bonding SB is higher than that of the LB.

With the amplitude $\delta u_B$ of the breathing mode,
the $E_\textrm{DFT}$ contribution is simply written as
$D^L_R\cdot \delta u_B= \varepsilon_\textrm{SB}-\varepsilon_\textrm{LB}$.
In case of a pristine octahedron, we can assign an overlap integral ($t_{sp}$) between $p$ orbital of oxygen and $s$ orbital of Bi(Sb) and the orbital energies for them as $\omega_p$ and $\omega_s$ respectively. 
Then, with distorted structures by the breathing mode, the energy splitting can be expressed up to the second order of $t_{sp}$ as follows,
\begin{equation}\label{eq:D_DFT_approx}
	D^L_R\delta u_B =\varepsilon_\textrm{SB}- \varepsilon_\textrm{LB} \sim \frac{4t^2_{sp}\delta \tau_{sp}}{|\omega_{p}-\omega_{s}|}, 
\end{equation}
where $t_{sp}$ changes to $t_{sp}(1\pm \delta\tau_{sp})$ for SB (LB) and 
$|\omega_{p}-\omega_{s}| \gg t_{sp}$ is assumed.
We note that the K doping ($x$) enhances the hole contribution to the $p$ orbital, thereby  
reducing the $|\omega_{p}-\omega_{s}|$. 
So, Eq.~\ref{eq:D_DFT_approx} implies that $D^L_R$ based on DFT-GGA increases with $x$, explaining the increasing matrix elements shown in Fig. 4. 
We also note that a small contraction of lattice constant with doping in Fig.~\ref{Fig.S.lattice} also could increases $t_{sp}$, thus corroborating the trend. 

With onsite and intersite Hubbard interaction for $s$ orbital of Bi(Sb) and $p$ orbital oxygen ($U_p$ and $V_{sp}$), the perturbative Hamiltonian for the Hubbard interactions can be written as 
\begin{equation}\label{eq:pert}
{\mathcal H}_\textrm{UV}={\mathcal H}_\textrm{U}+{\mathcal H}_\textrm{V}\equiv U_{p}(1-2n_{pp})-V_{sp}n_{sp},
\end{equation}
where $n_{pp}$ is the density matrix for the $p$ orbital of oxygen and $n_{sp}$ the generalized one for the neighboring $s$ orbital of Bi(Sb) and oxygen $p$ orbitals. Here, the spin index is suppressed and the significantly smaller onsite energy for $s$ orbital as shown in Table.~\ref{T2} is neglected.  
For $n_{pp}\neq 0$, the onsite Hubbard energy shifts the oxygen level of $\omega_p$ toward $\omega_s$ so that, according to Eq.~\ref{eq:D_DFT_approx}, the resulting $D^L_R$ should enhance compared with those from DFT-GGA
and increases as the amount of doped potassium accumulates. This tendency is indeed confirmed by our calculation only with $U$ as shown in Fig.~4.  

From Eq.~\ref{eq:pert}, we can estimate the contribution from the nonlocal interactions for the matrix element as follows,
\begin{equation}\label{eq:full_matrix}
D^L_R \delta u_B = \epsilon_\textrm{SB}-\epsilon_\textrm{LB}
\simeq \tilde{\varepsilon}_\textrm{SB}-\tilde{\varepsilon}_\textrm{LB}-V_{sp}(n_{sp}^\textrm{SB}-n_{sp}^\textrm{LB}),
\end{equation}
where $\tilde{\varepsilon}_\textrm{SB(LB)}$ is the renormalized energy level of Eq.~\ref{eq:D_DFT_approx} by the on-site energy of $U_p$ in Eq.~\ref{eq:pert} and
$n_{sp}^\textrm{SB(LB)}$ is the generalized density matrix for the SB (LB). Here we note that we neglect the perturbation to the wave function in considering Eq.~\ref{eq:full_matrix} that seems to be smaller than the variations of density matrix under perturbations.
The density matrix for the lower energy state should be larger than that for the higher, i.e.,  $n_{sp}^\textrm{LB}>n_{sp}^\textrm{SB}$ so that
Eq.~\ref{eq:full_matrix} immediately implies the enhanced $D^L_R$ with $V$ as shown in Fig. 4 of the main text. Furthermore, as doping increases, the charge disproportionation or difference between the SB and LB should decreases and then disappear as $x$ approaches 1. Therefore, $n_{sp}^\textrm{SB}-n_{sp}^\textrm{LB}$ in Eq.~\ref{eq:full_matrix} decreases to zero as doping increases so that the effect of intersite Hubbard interactions to $D^L_R$ diminishes as $x$ increases as shown in Fig. 4. 

\subsection{D. Comparison between DFT+$U$ and DFT+$U$+$V$}

\begin{figure}[t]
	\includegraphics[width=1.0\linewidth]{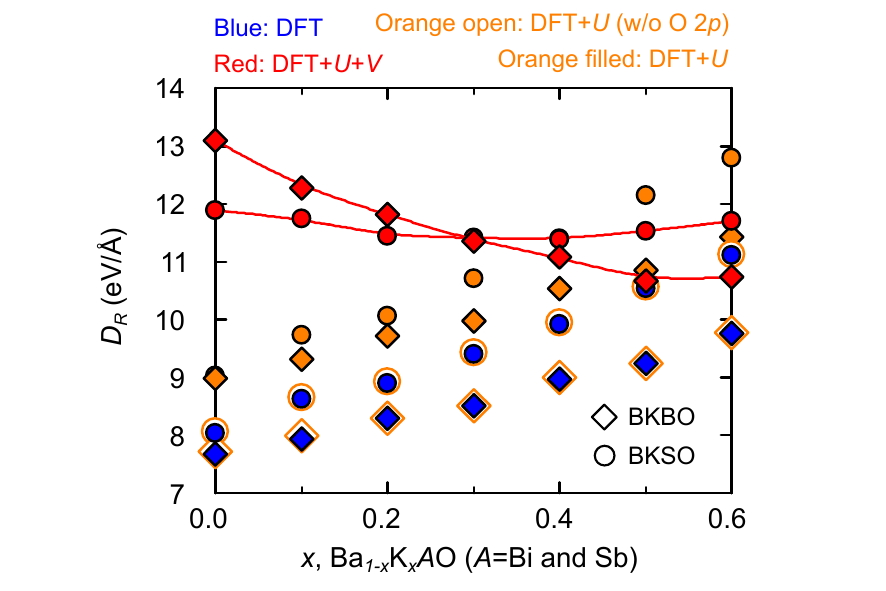}%
	\caption{Reduced $e$-$ph$ matrix element ($D^L_R$) as a function of doping $x$. Diamonds (circles) denote BKB(S)O and blue, orange open, orange filled and red color indicate DFT-GGA, DFT+$U$ ($U^{Bi}_s, U^{Bi}_p$), DFT+$U$ ($U^{Bi}_s, U^{Bi}_p$, and $U^{O}_{p}$)  and \UV results, respectively} \label{Fig.S.def}
\end{figure}

Like in the previous study~\cite{Korotin2012}, DFT+$U$ on Bi $s$ (and $p$) atomic orbital does not give any improvement compared to DFT-GGA method. The optimized crystal structure (Table~\ref{T1}), double well potential (Fig.~\ref{fig:FigS1}), phonon frequency as a function of K doping level (Fig.~\ref{fig:FigS3}), and reduced $e$-$ph$ matrix element (Fig.~\ref{Fig.S.def}) obtained from DFT+$U$ ($U^{Bi}_s, U^{Bi}_p$) [or labeled as DFT+$U$ (w/o O $2p$)] are consistent with those from DFT-GGA. 

When $U^{O}_p$ is included in DFT+$U$ calculation, it seemingly gives improvement for some physical properties. Unlike a typical partially filled case, the $U^{O}_p$ in BB(S)O changes the energetic position of O $p$ orbitals with respect to the Bi(Sb) $s$ orbitals as explained in the main text. The O $p$ level is shifted down so that the energy difference between bands for O $p$ orbitals and the Bi(Sb) $s$ orbital becomes smaller and the covalency between them is enhanced (Fig.~\ref{fig:Fig2}). As a result, the breathing distortion and the band gap are enhanced compared to DFT-GGA as shown in Table~\ref{T1} and Fig.~\ref{fig:FigS1}. 

However, it is still not enough to describe charge ordered phase of BB(S)O. 
The band gap of the monoclinic CDW phase obtained from DFT+$U$ ($U^{Bi}_s, U^{Bi}_p$, and $U^{O}_{p}$) is about 70\% of that from \UV and experimental observation.
The computed phonon stabilities also fail to explain experimental phase boundaries as shown in Fig.~\ref{fig:FigS3}.   
The most notable failure can be found in the reduced $e$-$ph$ matrix element ($D^L_R$) as shown in Fig.~\ref{Fig.S.def}. As already discussed in the previous section, the inclusion of $U^{O}_{p}$ affects the energy level and the covalency so that $D^L_R$ is enhanced compared to DFT-GGA results. However, $D^L_R$ keeps increasing as the K doping increases like DFT-GGA result. Without intersite Coulomb interaction $V$, one cannot explain the $e$-$ph$ coupling depending on K doping correctly as explained in the previous section. The decreasing trend of $D^L_R$ depending K doping can be captured only in \UV calculations and it well agrees with the previous HSE results.


\end{document}